\documentclass[letterpaper,aps,prd,twocolumn,preprintnumbers,showkeys,nofootinbib,nobibnotes]{revtex4}
\usepackage{epsfig}
\usepackage{dcolumn}
\usepackage{bm}

\def\map{{{WMAP} }}

\def\ga{~\mbox{\raisebox{-.6ex}{$\stackrel{>}{\sim}$}}~}

\begin{document}  
\preprint{CITA-2003-14}

\title{Suppressing the lower Multipoles in the CMB Anisotropies}

\author{Carlo R. Contaldi$^1$}\email{contaldi@cita.utoronto.ca}
\author{Marco Peloso$^1$}\email{peloso@cita.utoronto.ca}
\author{Lev Kofman$^1$}\email{kofman@cita.utoronto.ca}
\author{Andrei Linde$^2$}\email{alinde@stanford.edu}

\address{$^1$CITA,
University of Toronto, 60 St. George Street, Toronto, M5S 3H8, ON,
Canada\\
$^2$Department of Physics,
Stanford University, Stanford, CA 94305-4060,USA}

\begin{abstract} 
The Cosmic Microwave Background (CMB) anisotropy power on the largest
angular scales observed both by WMAP and COBE DMR appears to be lower
than the one predicted by the standard model of cosmology with almost
scale free primordial perturbations arising from a period of inflation
\cite{cobe,Bennett:2003bz,Spergel,Peiris}. One can either interpret
this as a manifestation of cosmic variance or as a physical effect
that requires an explanation. We discuss various mechanisms that could
be responsible for the suppression of such low $\ell$
multipoles. Features in the late time evolution of metric fluctuations
may do this via the integral Sachs-Wolfe effect. Another possibility
is a suppression of power at large scales in the primordial spectrum
induced by a fast rolling stage in the evolution of the inflaton field
at the beginning of the last 65 e-folds of inflation. We illustrate
this effect in a simple model of inflation and fit the resulting CMB
spectrum to the observed temperature-temperature (TT) power
spectrum. We find that the WMAP observations suggest a cutoff at
$k_c=4.9^{+1.3}_{-1.6}\times 10^{-4}$Mpc$^{-1}$ at 68\% confidence,
while only an upper limit of $k_c < 7.4\times 10^{-4}$Mpc$^{-1}$ at
95\%. Thus, although it improves the fit of the data, the presence of
a cutoff in power spectrum is only required at a level close to
$2\sigma$. This is obtained with a prior which corresponds to equal
distribution wrt $k_c$. We discuss how other choices (such as an equal
distribution wrt $\ln k_c$ which is natural in the context of
inflation) can affect the statistical interpretation.

\end{abstract} 
\date{March 31,  2003} 


\maketitle 
 
\section{Introduction} 
The most appealing cosmological scenario emerged in the mid 1980s as
the flat cold dark matter model with a nearly scale free initial power
spectrum, characterized by a gravitational potential $k^3 \vert
\Phi_k \vert^2 \equiv P_{\Phi}(k)=A_s k^{n_s-1}$, with $n_s\approx
1$. Besides the usual ingredients like baryon content $\Omega_B$
and a present expansion rate $H_0$, the model has only a single free
parameter, namely the amplitude of initial scalar perturbations $A_s$.
However, the wealth of observational results over the past decade has
forced the inclusion of an extra parameter in the form of an effective
cosmological constant $\Omega_{\Lambda}$.

Inflation plays a crucial role in explaining the flatness and
homogeneity required by the observations, as well as providing a theory of the
primordial spectrum of cosmological perturbations
\cite{cobe,Bennett:2003bz,Spergel,Peiris,toco,boom,maxima,dasi,cbi,vsa,acbar}. The
simplest single field models of inflation
\cite{Starobinsky:te,New,Chaotic} predict a flat universe with
primordial perturbations which slightly deviate from exact scale
invariance, $n_s \approx 0.95 \pm 0.03$ or so \cite{Mukh}, while more
elaborate models with more parameters can go beyond this
interval. For instance in the simplest models of hybrid inflation with
two fields one can have $n_s> 1$ \cite{Hybrid}.  A long period of
inflation (with more than 65 e-folds of expansion) generates a
huge patch which is much larger than the presently observable
universe.

CMB experiments manifest a striking agreement with this standard
cosmological model \cite{Spergel,Contaldi,boom}. However, an
interesting result of the \map data, which confirms earlier COBE DMR
observations, is a lower amount of power on the largest scales when
compared to that predicted by the standard $\Lambda$CDM models
\cite{Bennett:2003bz,Spergel,cobe}. Accurate Monte Carlo simulations
of the \map observations indicate that only 0.7\% of the realizations
of the models studied in \cite{Spergel} have less power than the
observed quadrupole. This number, which approximates the probability
of observing a smaller quadrupole than the one observed by WMAP in a
realization of standard $\Lambda$CDM model, should not be compared to
the posterior probability for the theoretical quadrupole to be equal
or higher to the one predicted by standard $\Lambda$CDM model, given
the small value measured by WMAP. The latter is about 5\% for
$\Lambda$CDM best fit quadrupoles (as can be easily estimated using a
$\chi^2$ distribution for the quadrupole), and the difference between
the two has been the source of some confusion in the literature
\cite{note}.~\footnote{The agreement between the low quadrupole
measured by COBE and CDM models was discussed in~\cite{bjk}. The
better agreement found in that work is due to the fact that CDM models
give a lower quadrupole than $\Lambda$CDM models, and that the value
observed by COBE for the quadrupole was slightly higher than the one
measured by WMAP. See also~\cite{gaz} for similar interpretations of
the WMAP data.}

The small posterior probability for $\Lambda$CDM model may still be
taken as a hint that some other model may provide a better fit to WMAP
data. One could dismiss this result by attributing it to cosmic
variance. We may simply live in a patch of the universe where the low
$\ell$ multipoles happens to be small, for no special reason. One may
even argue that with many independent measurements (many bins) it
would be very surprising if the results of at least one of these
measurements would not deviate strongly from theoretical
expectations. Indeed, WMAP shows several such peculiarities at various
values of $\ell$. Therefore it would be premature to interpret the
smallness of the low multipoles as a serious problem of the standard
model, particularly in light of its great success in explaining and
predicting major features of our universe in great detail. Indeed this
is the attitude adopted by Ref. \cite{Spergel} who mention the
``intriguing'' smallness of the low $\ell$ multipoles, but overall
interpret their results as providing a strong confirmation of the
standard inflationary paradigm. However, the deficit in power observed
first by COBE DMR and now by WMAP is an interesting result, and it is
tempting to look for possible explanations for such an effect
\cite{jingfang,Tegmark:2003ve,Uzan:2003nk,Efstathiou,Lewis,Linde:2003hc}. 
In
this {paper} we discuss some candidates for the mechanism by which the
low $\ell$ multipoles may be suppressed. In other words, we shall
investigate possible modifications of the standard model which can
give much better probability to have small power in low $\ell$
multipoles.

We distinguish between two possibilities in suppressing the low $\ell$
amplitudes. One is related to the physics of the inflationary phase in
the {\it early universe}. Theoretically we have significant freedom in
the design of inflationary potentials and, consequently, in the shape
of the primordial power spectra
\cite{Kofman:1986wm,Salopek:1988qh,Kofman:1989ed,Lidsey:1995np}.  As
we will show, one can also use the freedom in choosing initial
conditions at the onset of inflation. This also provides a mechanism
to affect the primordial power spectra. Both effects can be related
through the existence of a stage when the velocity of the scalar field
is not negligible, either during or prior to the observed $65$-e-fold
stage of inflation, and hence the perturbations cannot be described
using the usual slow--roll parameter approximation. The drawback of
this approach is clear; it is simply a tuning of parameters to obtain
a feature close to the present horizon scale.  A similar problem is
encountered in other attempts
\cite{jingfang,Tegmark:2003ve,Uzan:2003nk,Efstathiou,Lewis,Linde:2003hc}.

As a second option, the suppression could be due to {\it late
universe} physics under the influence of an effective cosmological
constant. This is probably the most interesting possibility, because
it relates the suppression of CMB anisotropy power at horizon scales
$\sim {H}^{-1}$ to the smallness of the cosmological constant, which
becomes dominant precisely at times $\sim {H}^{-1}\,$. The realization
of such a mechanism would then link two seemingly unrelated coincidence
problems.

This paper is organized as follows. In Sections~\ref{late}
and~\ref{early} we comment on late and early universe physics effects
as a cause for the suppression. In Section~\ref{model} we present a
simple example of an inflationary model used to obtain a spectrum with
a cutoff. In Section~\ref{cmb} we fit two separate cutoff spectra to
the \map observations in an attempt to constrain the scale of the
cutoff. We discuss our results in Section~\ref{discussion}.

\section{Late universe case}\label{late}

Observations of late--time acceleration and low power on largest
scales challenge our simplest theoretical expectations. In both cases
the effects become manifest on scales comparable to the present size
of the horizon, and at a time equal to the present age of the
universe. This is seen as a coincidence problem: ``Why now?''.

Indeed, CMB observations suggest that the spectrum of density
perturbations is nearly flat on scales even a few times smaller than the
present size of the observable universe. Therefore, if we were born
$10^9$ instead of $1.4 \times 10^{10}$ years after the big bang,
we would see CMB anisotropies that do not display
any suppression of low multipoles (unless we were living in an unusual
part of the universe due to cosmic variance). Similarly, if we were
living $10^9$ years after the big bang, the value of $\Omega_\Lambda$
would be less than 1\%.

It is thus tempting to search for a physical mechanism relating these
two problems. This may require a nontrivial modification of gravity at
the horizon scale. For example, it would be appealing to have a
mechanism that would screen not only the cosmological constant
\cite{Arkani-Hamed:2002fu}, but also inhomogeneities on the scale
$O(H^{-1}) \sim t$, on which the effects of the cosmological constant
become manifest. Another idea is related to possible nonperturbative
effects due to eternal inflation. For example, under certain
assumptions concerning the choice of measure in the theory of eternal
inflation, one can come to a paradoxical conclusion that we should
live in a center of a spherically symmetric distribution of matter
\cite{Linde:1994gy}.  This could lead to suppression of the
large-angle anisotropy without affecting the small-angle effects.

However, these possibilities are very speculative. As a simpler
alternative one may investigate whether the late time evolution of the
metric fluctuations $\Phi$ could be responsible for the suppression
via the Integral Sachs--Wolfe (ISW) effect.

The observed large--angle CMB temperature anisotropies are generated
by scalar
perturbations to the FRW geometry, which in the longitudinal gauge can
be written as
\begin{equation}\label{metric}
ds^2=a^2(\eta)\left( (1+2\Phi)d\eta^2-(1-2\Phi) d\vec x^2\right).
\end{equation}
The anisotropy has a contribution of $\frac{1}{3}\Phi$ sources at
the last scattering time hypersurface $\eta_r$. In addition, an
integral contribution arises, in the case where  time derivative $ \Phi'$ is
non-vanishing. Here $a(\eta)$ denotes the scale factor, $\eta$ the
conformal time, and a prime  denotes differentiation wrt $\eta$. Consider
a scale free primordial spectrum $\vert \Phi_k
\vert^2=\frac{A^2}{k^3}$. Then, for the CMB angular power spectrum
amplitudes we can adopt the result of \cite{KS85} 
\begin{equation}\label{power}
\langle \left(\frac{\Delta T}{T} \right)^2 \rangle_{\ell}=
\frac{2\ell+1}{\ell(\ell+1)}
\frac{A^2}{36 \pi^2} K^2_{\ell} \ ,
\end{equation}
where the coefficient $K^2_l$ is given by 
\begin{eqnarray}\label{coeff}
K_\ell^2&=&2 \ell(\ell+1) \int_0^{\infty} \frac{dk}{k} \Big[
 j_\ell(k\eta_0)  \nonumber\\ 
&&+6\int_{\eta_r}^{\eta_0} d\eta f'_k
j_\ell(k(\eta_0-\eta)) \Big]^2 \ ,
\end{eqnarray}
where $f'_k$ is $\Phi'_k$ normalized to $\frac{A}{k^{3/2}}$.
The flat CDM model gives $\Phi'=0$ and $K_\ell^2=1$ at large
scales. The flat $\Lambda$CDM model with pure cosmological 
constant $\Omega_{\Lambda} \sim 0.7$
gives instead $K_\ell^2 > 1$, amplifying the power above the
Sachs--Wolfe plateau \cite{KS85}.

However, the coefficients $K_\ell^2$ may be smaller than unity with
particular solutions for the late time evolution of $\Phi$. This can
be realized if the interference between the two terms in the
integral~(\ref{coeff}) is destructive, which in turn constrains the
spectrum and evolution of the metric perturbation $\Phi_k(\eta)$. This
may require special models of the effective cosmological constant, or the
addition of isocurvature fluctuations in the quintessence field~\cite{isw}.
We will address these possibilities in a future investigation.

\section{Early universe case}\label{early} 

Here we will describe modifications of the simplest inflationary models that
can account for the suppression of the low $\ell$
multipoles. As we will see, these modifications
have a common cause, which is the presence of a primordial or
intermediate regime where some of the slow roll approximations are
not valid (without necessarily meaning the interruption of inflation). This
results in a departure from the slow roll approximation 
when one evaluates the power spectrum originated during inflation.

\subsection{Changing the Potentials}

There have been many suggestions in the past on how to alter the
amplitude of density perturbations produced during inflation, see
e.g. \cite{Kofman:1986wm,Salopek:1988qh,Kofman:1989ed,Starobinsky:ts,
Lidsey:1995np,Hodges:bf}. The simplest idea is to fine-tune the shape
of the potential.  In general, this procedure is far from being
trivial.  The amplitude of metric perturbations at the horizon
crossing in the slow-roll approximation is given by
\begin{equation}\label{hyb_eqn1}
k^{3/2} \Phi_k =  \frac{V^{3/2}(\phi)}{\sqrt{6} V'(\phi)} \ ,
\end{equation}
where in the r.h.s. $\phi(t)$ is a function of $k$ through $k \sim
a(t)H(t)$, we use the units $M_p = (8\pi G)^{-1/2} =1$, and $V'$
denotes the derivative wrt $\phi$. This formula is convenient to
gain a heuristic insight of the problem.  Once we start bending the
potential, change occurs both in $V(\phi)$ and in $V'(\phi)$, and the
relation between $\phi$ and the momentum $k$ of perturbations changes
as well.

To do so, one does not have to use complicated potentials.  One can
use, for instance, the simple generic renormalizable potential
\begin{equation}\label{ren}
V(\phi)=\frac{1}{2}m^2 \phi^2 +\frac{1}{3}\sigma \phi^3+\frac{1}{4}
\lambda \phi^4\ .
\end{equation} 
By an appropriate choice  of parameters we  can generate significant
features -- dips or peaks --  in the spectrum,
as demonstrated in  \cite{Hodges:1989dw} with exact solutions of the
gauge invariant mode-by-mode  equations for fluctuations. 
Obviously, we must tune the parameters to place the dip of the 
power spectrum $\Phi_k$ around the present day cosmological
horizon.

The situation is even simpler in the hybrid inflation scenario, which
has more parameters by construction. The value of $V(\phi)$ is mainly
determined by the cosmological constant $V_0$ that does not change
significantly during inflation. In this case, in order to make the
amplitude of the potential smaller (greater) one should simply
increase (decrease) the slope of the potential.

Suppose, for example, that we have $\Phi \sim 10^{-5}$ (as we should)
 in the theory where the potential looks like $V_0 +\alpha\phi$ at low
 $\phi$, corresponding to the number of e-folds $N_e \lesssim
 55$. Suppose also that at larger $\phi$ the slope gradually increases
 and becomes 10 times greater on the scale corresponding to $60< N_e<
 65$. Then on this scale the amplitude of perturbations will become 10
 times smaller.  This should lead to a primordial spectrum $\Phi_k$
 with an amplitude which is strongly suppressed at low $k$, which will
 give us a desirable effect in the CMB anisotropy. Again, this
 requires some fine-tuning of the position of the place where the
 slope of the potential changes. An analogous spectrum was found in
 \cite{Leach:2000yw} through a numerical mode-by-mode computation in
 the model $V(\phi) \approx V_0 + \lambda\phi^4\,$.

Moreover, potentials of this type naturally appear in the simplest versions 
of hybrid inflation in supergravity (F-term inflation \cite{Copeland:1994vg}).
 If one considers N=1 supergravity   with a minimal {K\"ahler} potential and
 a very simple superpotential $W=  S (\kappa\bar\phi\phi - \mu^2)$, one finds a
 potential
of the following general form \cite{Linde:1997sj}:
\begin{equation}\label{hyb_eqn2}
V(\phi) \approx V_0(1 +\alpha \ln \phi) + \lambda\phi^4 +... \, .
\end{equation}

Note that the potential is steep at small and at large $\phi$, and relatively 
flat for some intermediate values of $\phi$. This leads to the cutoff of the
 spectrum at very short distances, as well as on the very large scale.
 A similar behavior  was advocated in 
\cite{Bennett:2003bz,Spergel,Peiris} based on the WMAP data and LSS$+L_{\alpha}$ data.
  For certain natural values of
 parameters the infrared  cutoff of the spectrum occurs on the scale
 of the horizon. This is exactly what one needs to explain the suppression 
of the low $\ell$ multipoles of CMB.

\subsection{Kinetic regime}

It is usually assumed that the inflaton scalar field is moving very
slowly at the beginning of inflation. This picture goes back to the
early days of inflationary paradigm, when the field was supposed to be
trapped in the minimum of the effective potential
\cite{Old,New}. Therefore the initial speed of the field was supposed
to be zero. In the language of density perturbations,
${\delta\rho\over \rho} \sim {V^{3/2}\over V'} \sim {H^2\over
\dot\phi}$, this implied that the density perturbations produced at
the beginning of inflation were very large, in complete agreement with
the usual statement that the spectrum of density perturbations in
simplest one-field inflationary models is red.

However, in chaotic inflation \cite{Chaotic} the initial speed of the
field $\phi$ can be quite large.  The field eventually slows down when
it approaches the inflationary regime, which is an attractor in the
phase space of all possible trajectories  $(\phi(t), \dot\phi(t))$.
  For a small subset of initial conditions the field may
approach the inflationary trajectory relatively late, so that at the
onset of the last 65 e-folds of inflation the field will have higher speed than is usually expected. 
In this case the
amplitude of metric perturbations at the scale of the horizon will be
smaller than expected, which in turn will result in a lower variance
of CMB anisotropies at small $\ell$.
This marginal inflation requires a tuning of the initial conditions,
and not of the shape of the potential. 

A  similar situation  may  occur 
 in hybrid  inflation. The
simplest potential for two-field hybrid inflation is \cite{Hybrid}
\begin{equation}\label{hyb_eqn3}
V(\phi,\sigma) = {\lambda \over 4} (\sigma^2 - v^2)^2 + {g^2 \over 2}
\phi^2 \sigma^2 + {1 \over 2} m^2 \phi^2.
\end{equation}
The point where $\phi=\phi_c = {\sqrt{\lambda} \over g} v $ and
$\sigma=0$ is a bifurcation point.  For $\phi > \phi_c$ the squares of
the effective masses of both fields $m_\sigma^2 = g^2 \phi^2 - \lambda
v^2 + 3 \lambda \sigma^2$ and $m^2_{\phi}=m^2+g^2\sigma^2$ are
positive and the potential has a minimum at $\sigma=0$.  For $\phi <
\phi_c$ the potential has a maximum at $\sigma=0$.  The global minimum
is located at $\phi=0$ and $|\sigma|= v$.  Inflation in this model
occurs while the $\phi$ field rolls slowly from large values toward
the bifurcation point.

The space of initial conditions now
 is $(\phi(t), \dot\phi(t); \sigma(t), \dot \sigma(t))$.
In general evolution begins with large values of $\phi$ and $\sigma$, at the boundary with Planck energy density. 
At large $\phi$ the dominant contribution to $V$ is given by the term
$g^2\phi^2\sigma^2/2$, so the Planck boundary is given by a set of
four hyperbole \cite{NO}
\begin{equation}\label{222}
 g |\phi ||\sigma| \sim 1  \ .
\end{equation}
We will assume that initially $|\phi|> |\sigma|$ and take $g \ll 1$.
The field $\phi$ on this branch can take any value up to $g^{-1}$ (for
greater values of $\phi$ the field $\sigma$ would have a
super-Planckian mass $g\phi$). Suppose for definiteness that $|\phi |$
is close to its upper bound, so that the initial value of the field
$\sigma$ at the Planck boundary is of order $1$ or somewhat
greater. This allows for a short stage of chaotic inflation supported
by the potential of the field $\sigma$ when this field rolls down
toward $\sigma = 0$.

After the first stage of inflation, the field $\sigma$ rapidly
oscillates, with the frequency just slightly smaller than the Planck
mass, and with the amplitude $\sigma(t)$ decreasing as $a^{-3/2}$. The
oscillations induce a large contribution to the effective mass of the
field $\phi$: $m^2_\phi = m^2 + g^2 \langle\sigma^2\rangle$. Here
$\langle\sigma^2\rangle \sim a^{-3}$ stays either for the average
square of the amplitude of the oscillations of the field, or for the
fluctuations of this field produced during its decay to particles
$\phi$ and $\sigma$.

In the beginning, $g^2 \langle\sigma^2\rangle \gg m^2$, so the field
moves with a much greater speed than what one could naively expect by
looking at the effective potential of the field $\phi$ and ignoring
its interactions with the field $\sigma$ \cite{NO,Cline:2003gq}. This
leads to a strong suppression of density perturbations produced at the
first stages of hybrid inflation. However, eventually the term $g^2
\langle\sigma^2\rangle$ drops down and we enter the standard hybrid
inflation regime producing perturbations with nearly flat
spectrum. Since the term $g^2\langle\sigma^2\rangle$ drops down very
rapidly, as $a^{-3}(t)$, the transition between the flat spectrum
produced at late stages of inflation and the strongly suppressed
spectrum at the early stages of inflation occurs very abruptly, within
a single e-fold of inflation. This is exactly what we want in order to
explain the absence of the low $\ell$ multipoles.

In terms of the slow roll parameters $\epsilon=\frac{\dot
  \phi^2}{2H^2}$, $\delta=-\frac{\ddot \phi}{H\dot \phi}$, one has
  $n_s-1=-2\delta-4\epsilon$. \footnote{Note that these parameters can
  be further expressed in terms of the slow--roll parameters
  $\epsilon$, $\eta$, $\lambda$, etc., given in terms the inflaton
  potential and its derivatives. For our purposes however, it is
  convenient to keep the original parameters given in the text.}
  Suppression of the spectrum at small $k$ implies that $n_s-1$, is
  large for $k$ corresponding to the scale of the present horizon.
  This means that in our examples some of the slow-roll parameters
  $\epsilon, \delta$, etc., at some early (or intermediate) stage
  of inflation were large. This can be attributed to the time
  derivatives $\dot \phi$, $\ddot \phi$, etc.  Thus, departure from
  the slow roll approximation for calculating power spectrum is
  related to the regime where kinetic terms cannot be neglected.

We would like to mention also, that there is a new possibility of
modulated cosmological perturbations from inflation \cite{mod1,mod2},
where one may use a different type of tuning to get desirable spectra
from inflation, while leaving the inflaton potential unchanged.

\section{Illustration of the cutoff spectrum}\label{model}

For illustrative purposes, we now study in a more detailed way one of
the simplest possibilities we have discussed above, namely the
existence of a period of fast rolling of the inflaton field $\phi$
(with strong kinetic domination) before the last stage of
inflation. The period of inflation is assumed to last for about
$60-65$ e-folds (the exact length depending on the details of
reheating) such that the stage of fast roll can leave an imprint
precisely at the largest scales we presently observe. As a working
assumption, we start with a homogeneous and flat universe already at
the stage of fast roll, so that one may have to postulate the
existence of a previous period of inflation at earlier times.  Apart
for this requirement, we leave the details of the earlier universe
unspecified, and we rather concentrate on the evolution of the
background and of the cosmological perturbations starting from the
fast roll regime.

We are mainly interested in the spectrum of primordial
perturbations. We first perform an exact mode-by-mode
numerical computation in the
context of chaotic inflationary potential and large initial velocity
for the inflaton field. We will then compute the spectrum in a simpler
idealized situation, with an instantaneous transition between a regime
of kinetic domination and a nearly de-Sitter stage. In this last case
one can derive a simple analytical result which reproduces very well
the exact spectrum obtained in the numerical evolution.

We start with a quadratic inflaton potential, $V \left( \phi \right) =
m_\phi^2 \, \phi^2 /2\,$, and initial conditions $\phi_{\rm in} = 18.0
\, M_p\;,\; \left( d \phi/ d t \right)_{\rm in} \simeq - 10\,
m_\phi\,\phi_{\rm in}\,$, giving about $61\,$ e-folds of inflation.
Concerning the perturbations, we work in momentum space and focus on
the Mukhanov variable \cite{muk}
\begin{equation}
v_k \equiv a \left( \delta \phi_k + \frac{\phi'}{h} \, \Phi_k \right)
\end{equation}
where $\delta \phi_k$ denotes the perturbations in the inflaton field,
$\Phi$ is the metric perturbation given in Eq.~(\ref{metric}), 
prime denotes derivative with respect to conformal time $\eta$,
 and $h
\equiv a'/a\,$, $a$ being the scale factor of the universe. The
evolution of $v_k$ is given by (see e.g. \cite{mfb} for details)
\begin{equation}
v'' + \left[ k^2 - \frac{z''}{z} \right] v = 0 \quad,\quad\quad
z \equiv \frac{a \phi'}{h}
\label{eqv}
\end{equation}

Due to the choice of initial conditions, the system is initially in a
kinetic dominated regime, with $h \propto \phi' \,$. As a consequence,
we have initially
\begin{eqnarray}
a &\simeq& \sqrt{1+2 \, h_0 \, \eta} \nonumber\\
\frac{z''}{z} &\simeq& \frac{a''}{a}
\simeq \frac{- \, h_0^2}{\left(1 + 2 \, h_0 \eta \right)^2}
\label{zsz}
\end{eqnarray}
with $h_0$ denoting the value of $h$ at the time $\eta=0$, where we
have normalized $a = 1\,$.

As long as the approximation (\ref{zsz}) is valid, Eq.~(\ref{eqv}) is 
solved by
\begin{equation}
v_k \left( \eta \right) = \sqrt{\frac{\pi}{8\,h_0}} \,
\sqrt{1 + 2 \, h_0 \, \eta} \, H_0^{(2)} \left( k \, \eta +
\frac{k}{2\,h_0} \right)
\label{solv}
\end{equation}
where $H_0^{(2)}$ is a specific Hankel's function and where we have
chosen (and appropriately normalized) the solution of Eq.~(\ref{eqv})
which reduces to $\,{\rm exp} \left( - i \, k \, \eta \right)/ \sqrt{2
\, k}\,$ at very short wavelengths, $k \gg 1/ \vert \eta \vert
\,$.~\footnote{The choice~(\ref{solv}) can be motivated by
taking an adiabatic vacuum during a previous inflationary stage, and
assuming an adiabatic transition into the kinetic dominated regime
(adiabaticity may be preserved due to a high effective mass for the
field during these previous times, as for example in the hybrid
inflationary case described in the previous section). An analogous
initial state was considered for example in \cite{sahni}, where the
production of gravitational waves was studied in a model with a
radiation dominated stage followed by inflationary expansion.}

In the numerical integration we have solved exactly the evolution
equations for the background quantities $\phi$ and $a\,$, together
with Eq.~(\ref{eqv}) for the perturbation mode $v_k\,$. We have used
Eq.~(\ref{solv}) only to set the value of $v_k$ and its derivative at
the initial time $\eta=0\,$. We have plotted in the last panel of
fig.~\ref{fig:fig2} the final spectrum for the metric perturbation 
$\Phi_k\,$.

Modes with very large $k$ are well inside the horizon during the fast
roll of $\phi\,$, and hence they are quite insensitive to the
evolution of the background at this stage. Their horizon crossing
occurs during inflation, and the usual computation (giving nearly
scale invariant spectrum) applies.  At wavelengths comparable with the
size of the horizon $H_*^{-1}$ during the onset of inflation we find
instead a strong departure from scale invariance.  The amplitude of
the spectrum is oscillating for $k \ga H_*$, followed by a sharp
decrease at $k \sim H_*\,$. By an appropriate choice of the number of
e-folds of inflation, it is possible to relate such wavelengths to the
size of the present horizon, with a consequent decrease of the
observed CMB anisotropies at large scales.

This result is well reproduced in a highly simplified situation, with
an instantaneous transition between the regimes of kinetic
domination and nearly de-Sitter expansion. The scale factor $a$ in the
idealized model evolves as
\begin{eqnarray}
a \simeq \sqrt{1+2\,H\,\eta} &,&\quad \eta \leq 0 \nonumber\\
a \simeq \frac{1}{1-H\,\eta} \quad &,&\quad \eta \geq 0
\label{scale}
\end{eqnarray}
where we have now set $\eta = 0\,$, and $a=1$ at the transition, while
$H$ denotes the (physical) Hubble parameter during inflation.

The evolution equation for the Mukhanov variable $v_k$ is again given
by Eq.~(\ref{eqv}), with $z''/z \simeq a''/a$ in both regimes. During
kinetic domination, the evolution of $v_k$ is again given by
Eq.~(\ref{solv}) with the quantity $h_0$ replaced by $H\,$. During the
de-Sitter stage one finds instead
\begin{eqnarray}
v &=& C \, {\rm e}^{-i \left( k \, \eta - k/H \right)}
\left( 1 - \frac{i}{k \, \eta - k/H} \right) + \nonumber\\
&+& D \, {\rm e}^{i \left( k \, \eta - k/H \right)}
\left( 1 + \frac{i}{k \, \eta - k/H} \right)
\label{latesol}
\end{eqnarray}
and the coefficients of the two modes can be obtained by requiring
continuity of $v$ and $v'$ at the transition,
\begin{eqnarray}
C &=& \frac{{\rm e}^{-i\,k/H}}{\sqrt{32\,H/\pi}}
\,\left[ H_0^{(2)} \left( \frac{k}{2\,H} \right) -
\left( \frac{H}{k} + i \right) \,
H_1^{(2)} \left( \frac{k}{2\,H} \right) \right] \nonumber\\
D &=& \frac{{\rm e}^{i\,k/H}}{\sqrt{32\,H/\pi}}
\, \left[ H_0^{(2)} \left( \frac{k}{2\,H} \right) -
\left( \frac{H}{k} - i \right) \,
H_1^{(2)} \left( \frac{k}{2\,H} \right) \right] \nonumber\\
\label{cd}
\end{eqnarray}

We consider the spectrum of $Q \equiv v/a$ which becomes
constant at late time, as indicated by the leading order
contribution to equation~(\ref{latesol}) for $\eta
\rightarrow 1/H\,$
\begin{equation}
P_Q \equiv \frac{k^3}{2 \, \pi^2} \vert Q \vert^2 \rightarrow
\frac{H^2}{2\,\pi^2} \, k \, \vert C - D \vert^2
\label{spect}
\end{equation}
At short wavelengths ($k \gg H$) one recovers the standard
result of a (nearly) scale invariant spectrum
\begin{eqnarray}
\vert C \vert &\simeq& \frac{1}{\sqrt{2 \, k}}
\quad,\quad \vert D \vert \ll \vert C \vert \nonumber\\
P_Q  &\simeq& \left( \frac{H}{2\,\pi} \right)^2
\label{largek}
\end{eqnarray}

At super-horizon scales, the two modes $Q_k$ and $\Phi_k$ are related
by a $k$ independent rescaling so that the spectrum given by
Eqs.~(\ref{cd}) and (\ref{spect}) directly translates into the
spectrum of $\Phi\,$, up to an overall normalization factor. 

\begin{figure}[t]
\centerline{\psfig{file=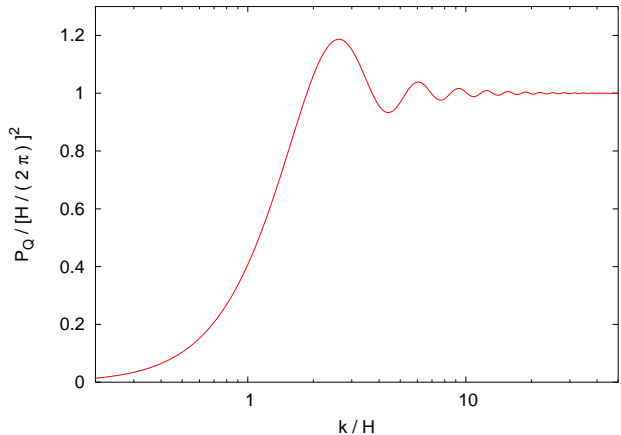,width=6 cm,angle=-90}}
\caption{\label{fig:fig1} Power spectrum for $Q$ from the analytical
computation, Eqs.~(\ref{cd}) and (\ref{spect}). This is to be compared
with the spectrum of $\Phi$ obtained by the numerical evolution, and
reported in the last panel of fig.~\ref{fig:fig2}.}
\end{figure}

The spectrum obtained by this simple analytical calculation is shown
in figure~\ref{fig:fig1}. We see that it reproduces very well the
spectrum of $\Phi$ obtained with the exact numerical evolution, which
is reported in the last panel of fig.~\ref{fig:fig2}.

\section{CMB anisotropies with a cutoff}\label{cmb}

To determine how strongly the data favour a primordial power spectrum with a
feature at large scales we have fit the \map data with two template
cutoffs. The first is derived from the simple model discussed
above and includes the detailed oscillatory effect at the transition
between the two regimes in the initial spectrum. The second is based
on an exponential cutoff of the form
\begin{equation}\label{cutoff}
  k^3|\Phi(k)|^2 = A_s (1-\exp^{-(k/k_c)^{\alpha}}) k^{n_s-1},
\end{equation}
where we have chosen $\alpha$ to fit the rise of the model with
$\alpha=3.35$. $A_s$ is the overall amplitude of the scalar
perturbations. This parametrization does not include features in the
power spectrum which are dependent on the specific model considered,
but which are too small to change qualitative features in the CMB
power spectrum.

Due to the nature of the TE cross-correlation signal reported by \map,
we chose to fit our models to the TT data only. We believe that it is
still premature to draw too strong conclusions by making use of the
E-type polarization signal, which has only been measured through its
correlation with the much larger temperature signal by WMAP. Once the
\map E-type observations have reached a sufficiently high
signal--to--noise level to unambiguously measure a polarization signal
free of systematics, its use in constraining models will be much
safer.~\footnote{An analysis with inclusion of TE data was first
performed in~\cite{Lewis}, and, successively to~\cite{Lewis} and to
the first version of the present manuscript, in ~\cite{cline}. The
results of~\cite{cline} are similar to ours, but their conclusions
concerning the problem of low $\ell$ multiples are slightly more
optimistic due to the strong emphasis put on the TE data. We caution
that taking into account both cosmic variance and the correlation with
the much larger TT signal would result in a significant increase of
the error bars on the TE data (see fig. 2 of ref.~\cite{Lewis}, where
this is explicitly shown for a particular model), possibly affecting
claims which strongly depend on their use.}

We use a modified version of the publicly available CMBFAST
\cite{cmbfast} code to compute CMB spectra with various $k_c$ in the
two cutoffs. We generate grids of models by varying three parameters
which affect the shape of the spectrum at the lowest multipoles,
namely the spectral index of the primordial power spectrum $n_s$, the
energy density of cosmological constant component in units of the
critical density $\Omega_{\Lambda}$, and the present day wavenumber of
the cutoff in the spectrum $k_c$ in units of Mpc$^{-1}$. The grid
values are regular in the parameters and have a resolution of $32,20$,
and $32$ grid points with ranges $[0.83,1.04]$, $[0.6,0.85]$, and
$[2.0\times10^{-3},1.0\times10^{-5}]$ respectively.

At each point in the grid we use subroutines derived from those made
available by the \map team to evaluate the log likelihood with
respect to the \map data \cite{Verde}. Other parameters that are not
expected to affect the shape of the spectrum at $\ell < 30$ are held
fixed at the following values; $\Omega_bh^2=0.022$, $\Omega_{m}h^2 =
0.135$ and $\tau_c=0.17$ \cite{Spergel}. In addition we consider only
flat models with $\Omega_{\rm tot}=1$. The present value of the Hubble
constant for each model is then fixed for each choice of
$\Omega_{\Lambda}$ in the grid. As we are not interested in the
overall amplitude of the primordial perturbations we marginalize over
a range of values in the amplitude of the power spectrum. We only
consider scalar perturbations in this work.

Our choice of parameters is motivated by our interest in motivating
the requirement for a cutoff in the spectrum and not by an attempt at
a precise determination of cosmological parameters, an exercise which
should be the focus of much more exhaustive investigations
\cite{Spergel,Lewis}. Residual correlations with parameters that have
been fixed such as $\tau_c$ will be subsumed by our tilt and amplitude
parameters which should be viewed as mildly {\it effective}
quantities. In addition our use of a prior on $\Omega_{\Lambda}$ (see
below) is motivated by a simple mimicking of Large Scale Structure
(LSS) and Type--Ia supernovae (SN1a) constraints external to the \map
results. \footnote{The authors of \cite{cline} erroneously compare our
analysis to their ``\map TT only'' results. More appropriately, our
simple choice of parameters and priors is closer to a ``WMAPext TT +
2dF + SN1a'' combination. The latter requires a very different
parameter grid compared to the ``\map TT only'' data, whose best fit
model lies out in the tail of the LSS+SN1a motivated priors
\cite{Spergel}.}

\begin{figure}[t]
\centerline{\psfig{file=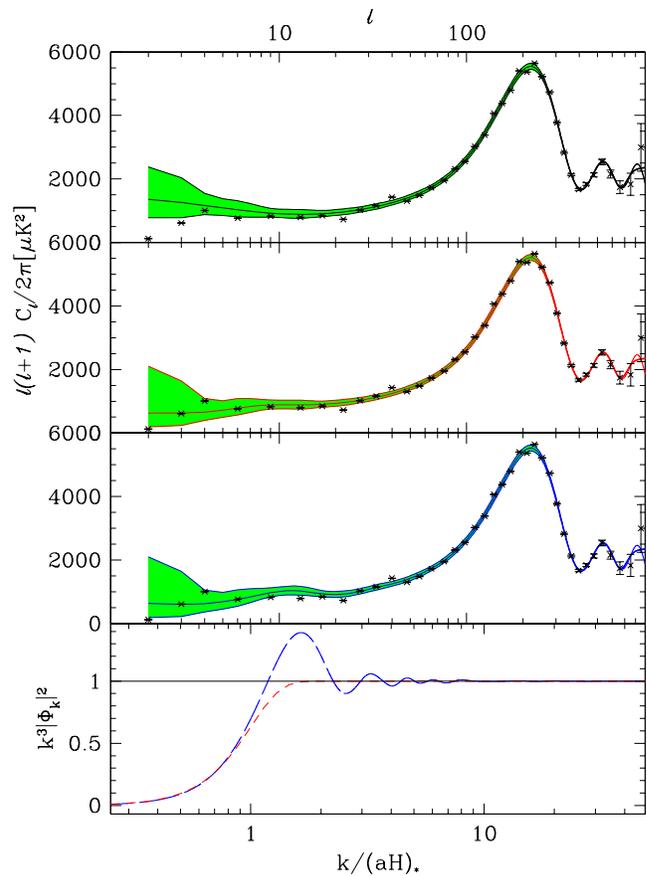,width=9 cm,angle=0}} 
\caption{\label{fig:fig2} CMB anisotropies with cutoff primordial
  spectra. The top panel shows a standard $\Omega_{\Lambda}CDM$
  model without cutoff, and with $n_s=0.9587$ and
  $\Omega_{\Lambda}=0.7316$, corresponding to the best--fit values of
  our model grids. The upper and lower contour show the 1$\sigma$
  confidence level for a lognormal distribution appropriate for the
  sample variance of the observations. The error bars show the
  experimental noise contributions. The result in the second panel is
  obtained with the same model, but with an infrared cutoff on the
  primordial power spectrum, as obtained from an exact evolution of the
  model described in section~\ref{model}. The result in the third panel
  is instead obtained with the power spectrum given in Eq.~(\ref{cutoff}).
  The two cutoffs chosen correspond to the best--fit cutoff scales
  $k_c=4.9\times 10^{-4}$Mpc$^{-1}$ and $k_c=5.3\times
  10^{-4}$Mpc$^{-1}$ respectively. The bottom panel shows the two
  primordial power spectra corresponding to these cutoffs and
  for the case $n_s=1$.}
\end{figure}

In Fig.~\ref{fig:fig2} we show the resulting best--fit models we
obtain from the grid for both cutoff models considered. The top panel
shows a standard (no cutoff) spectrum model with the same
parameters as those obtained in the best--fit for the two models
($n_s=0.9587$ and $\Omega_{\Lambda}=0.732$). We have displayed the
noise contribution to the uncertainties as errors on the binned \map
results. The cosmic variance contribution to the errors is shown as
upper and lower contours around the model. We have displayed the
$1\sigma$ confidence limits for a lognormal distribution as this
approximates more clearly the significance of the low quadrupole and
octupole signal. The middle two panels show the best--fit for the
model cutoffs. We can see that both models fit the lower multipoles
better then the standard spectrum by reducing power
on the largest scales. The bottom panel shows the two template spectra
used in the fits shifted to an arbitrary cutoff scale of $k_c=1$ for a
$n_s=1$ model.

The bottom panel shows the two template spectra used in the fits shifted
to an arbitrary cutoff scale of $k_c=1$ for a $n_s=1$ model. It is
important to note that even a step--like cutoff in the initial spectrum
will not produce a step--like feature in the CMB spectrum due to the
convolution of the initial spectrum in Eq.~\ref{coeff}. Thus it is
difficult to reproduce the sharp drop--off that occurs in the angular
power spectrum.

Although we fit for all TT multipoles in the range $2\le\ell\le900$,
most of these simply provide an accurate calibration of the overall
amplitude of the model as we are keeping the matter content fixed. We
therefore concentrate on the lowest multipoles in considering the
significance of our best--fit models. For $\ell\le 10$ we find that
the probability of exceeding the observed $\chi^2$ is 5.6\% in the
case of the standard model, 21.7\% when using the template model
cutoff, and 33.4\% with the exponential cutoff of Eq.~\ref{cutoff}.

\begin{figure}[t]
\centerline{\psfig{file=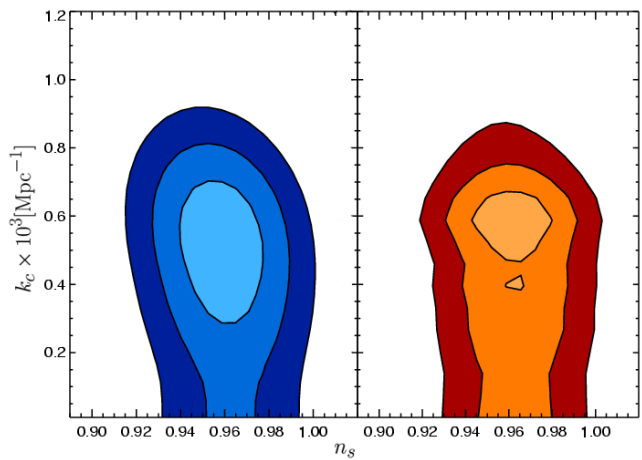,width=8.5 cm,angle=0}} 
\caption{\label{fig:fig3} Marginalized distribution in the
  $n_s$--$k_c$ plane for our grid of models. The left panel is for the
  exponential cutoff while the right panel is for the toy model
  discussed in the text. The solid contours show drops
  in $\chi^2$ corresponding to 68\%, 95\%, and 99.7\% confidence for a
  two--dimensional Gaussian distribution.}
\end{figure}

In Fig.~\ref{fig:fig3} we marginalize over the range in
$\Omega_{\Lambda}$ with a Gaussian prior $\Omega_{\Lambda}=0.70\pm
0.03$ and plot the resulting $\chi^2$ values as a function of $n_s$
and cutoff scale $k_c$. The contours shown are for $\Delta\chi^2$
values giving one, two, and three $\sigma$ contours for two parameter
Gaussian distributions. We find that the distribution for the cutoff
scale has a tail as $k_c\rightarrow 0$ close to the $2\sigma$ level.
This is evident from the fact that the models asymptotically approach
the scale invariant case which has a low but not vanishing probability
with respect to the observations \cite{Spergel}. The data however
prefer a cutoff scale within the present observable universe
at the $1\sigma$ level.

\begin{figure}[t]
\centerline{\psfig{file=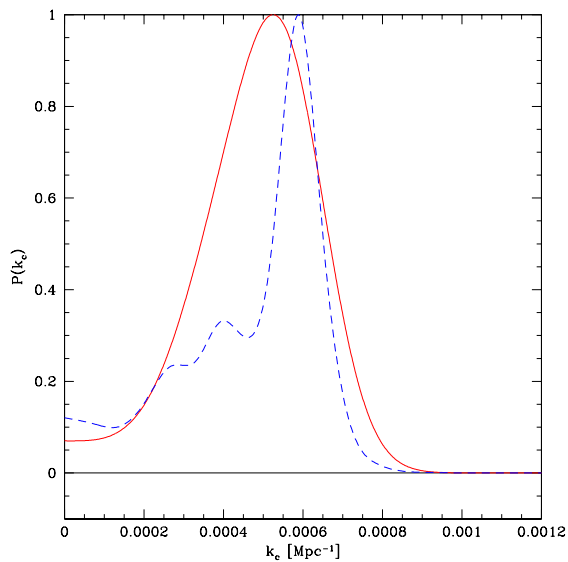,width=8 cm,angle=0}} 
\caption{\label{fig:fig4} The one dimensional marginalized
  distribution for the cutoff scale $k_c$. The solid (red) curve is
  for the exponential cutoff while the dashed (blue) curve is for the
  spectrum obtained in the inflation model discussed in the
  text. Notice that the sharp drop--off in the likelihood at high $k_c$
  is driven mainly by our prior on $\Omega_\Lambda$.}
\end{figure}

Finally we marginalize over $n_s$ with a flat prior to obtain the
marginalized probability distribution in $k_c$ shown in
Fig.~\ref{fig:fig4}. As the distributions do not vanish as
$k_c\rightarrow 0$ we integrate the functions in the range $0\le
k_c\le \infty$ to obtain 68\% and 95\% confidence limits. For the
inflation model we find $k_c=5.3^{+0.9}_{-2.6}\times
10^{-4}$Mpc$^{-1}$ with a 95\% upper limit of $k_c < 7.02\times
10^{-4}$Mpc$^{-1}$. Similarly for the exponential cutoff model we find
$k_c=4.9^{+1.3}_{-1.6}\times 10^{-4}$Mpc$^{-1}$ with a 95\% upper
limit of $k_c < 7.4\times 10^{-4}$Mpc$^{-1}$. It is important to note
that the confidence limits depend on the integration measure
adopted. In this case we have taken a measure linear in $k_c$; however
an alternative would be to take a measure in $\ln k_c$ which may be
more suitable for the distribution of the cutoff scales given by
inflation theories. This second option, which gives higher weights to
lower wavenumbers, will crucially depend on the value chosen for the
largest scales, $\ln k_c^{\rm min} > -\infty$. In the context of
inflationary models it is natural to limit to those scales which left
the horizon when the energy density of the universe was close to
Planckian.

When faced with such ambiguities it is useful to refer to the
difference in likelihoods between different values of $k_c$. This
allows an approximate estimate of the significance of having a cutoff
without depending on any measure used in deriving confidence limits.
As an example the difference in $\chi^2$ between the peak in the
distributions and at $k_c=0$ is found to be $\Delta\chi^2=5.3$ for the
exponential cutoff and $\Delta\chi^2=4.2$ for the inflation
model. This indicates that the case with no cutoff is just over
$2\sigma$ below our best--fit cutoff model.

\section{Discussion}\label{discussion}

Both COBE DMR and \map observations indicate a low value for the power
on largest scales. Interestingly, these scales are of the same order
as the scale corresponding to the time of dominance of the
cosmological constant.  If one ignores the possible interpretation of
the suppression of low multipoles due to cosmic variance (which is
perhaps the simplest and the least dramatic interpretation), one faces
a new coincidence problem in addition to the usual one required to
explain the late--time acceleration of the universe
(i.e. $\Omega_{\Lambda}\sim 2\Omega_m$ today).  In this paper we
briefly discussed the possibility that the late--time evolution of the
metric perturbations $\Phi_k(t)$ may reduce the power in the low
multipoles through integrated Sachs--Wolfe effect. This may require a
combined investigation of adiabatic and isocurvature perturbations in
certain models of dark energy.

The main emphasis of our paper was related to modifications of the
simplest models of inflation due to the possible existence of a
kinetic stage, when the velocity of the scalar field was not
negligible. The breaking of the slow roll approximation which could
occur at this stage may cause a significant suppression of the large
scale density perturbations. In order to affect the low $\ell$
multipoles, this stage should occur in the beginning of the last $65$
e-fold period of inflation. Clearly, this requires either fine tuning
of parameters in the inflaton potential, or fine-tuning of initial
conditions which are necessary to obtain a desirable modification of
the spectrum close to the present horizon scale.

The final part of the paper has the aim to quantify with a simple
example the level of suppression required by the data. A stage of
kinetic domination before the observable stage of inflation can result
in a sharp cutoff in the spectrum of primordial perturbations. 

Fits to the observations reveal that the data favour a cutoff in the
spectrum although the significance of the ``detection'' is less than
$3\sigma$. This is a reflection of the fact that the $k_c=0$ case (i.e.
standard $\Lambda$CDM model) is not a terrible fit to the data. The price
of obtaining a better fit is the addition of an extra parameter in the
theory which should probably be done under stronger observational evidence
than we currently have. 

On the other hand, we should note that when we attempt to evaluate all
possible theories that could produce the required suppression of the
spectrum, we consider the space of all possible cutoff parameters
equally distributed in momentum space. Meanwhile if we would assume
equal prior distribution in space of parameters of inflationary
models, this would translate into an approximately equal distribution
in terms of $\ln k$ rather than $k$.  This remark also applies to
other attempts to explain the suppression of low multipoles
\cite{jingfang,Tegmark:2003ve,Uzan:2003nk,Efstathiou} and constitutes a 
choice
of prior on $k_c$ a problem encountered when constraining many of the
cosmological parameters (e.g. a choice of uniform prior on $H_0$ as
opposed to in $\Omega_\Lambda$). However, as we have discussed, a
possibly less ambiguous quantity is the likelihood ratio between
different values of the cutoff since this does not depend on any
underlying theoretical distribution. We have quantified this in the
previous section.

We should also bear in mind that in any case, observations tell us the
values of the parameters, and we should accept them even if they seem
unusually fine--tuned.  For example, many coupling constants in the
standard theory of electroweak interactions are of the order
$10^{-1}$. Meanwhile the coupling constant of the electron and the
Higgs field, which is responsible for the electron mass, is $2\times
10^{-6}$. The smallness of this parameter is notable, but
usually we do not assume that all coupling constants are random
variables and do not calculate the probability to have such a small
coupling constant on the basis of its comparison to other coupling
constants in the theory. We still do not know why this coupling
constant is so incredibly small, but it does not make the standard
theory of electroweak interactions unacceptable. Moreover, according
to the standard model, the neutrino must be massless, meanwhile the
recent data suggest that this is not the case. This means that we need
to extend the standard model by making it even more
complicated. However, this does not cast any doubts concerning the
basic idea of gauge invariance and spontaneous symmetry breaking,
which made the electroweak theory internally consistent.

High energy physicists never claim that they fine-tune the parameters
of the theory, they simply fit the data. Until very recently
cosmologists did not need to do the same and had the luxury to debate
which values of the parameters seemed more natural. We are now entering
the age of precision cosmology, and we need to adjust our attitude
accordingly.

During the last 20 years, the inflationary scenario emerged as the
simplest and possibly even unique way of constructing an internally
consistent cosmological theory. The main result of our paper is that,
if needed, we can fit the parameters of the inflationary theory and
make it consistent with the small amplitude of the large-angle
anisotropy of the CMB. It is incomparably easier to do so than to
explain all features of the spectrum of the CMB, as well as the
observed homogeneity, isotropy and flatness of the universe without
using inflation.

\acknowledgments
 
It is a pleasure to thank Dick Bond, Juan Garcia-Bellido, Mark
Halpern, Johannes Martin, Dmitri Pogosyan, David Spergel, Max Tegmark,
and David Wands for helpful discussions. We particularly thank Antony
Lewis for detailed discussions on his related work
\cite{Lewis}. Research at CITA is supported by NSERC and the Canadian
Institute for Advanced Research. The computational facilities at CITA
are funded by the Canadian Fund for Innovation.  The work by A.L. was
also supported by NSF grant PHY-9870115 and by the Templeton
Foundation grant No. 938-COS273.


\begin{thebibliography}{999} 

\bibitem{cobe}  C.~L.~Bennett {\it et al.},
``4-Year COBE DMR Cosmic Microwave Background Observations: Maps and Basic Results,''
Astrophys.\ J.\  {\bf 464}, L1 (1996)
[arXiv:astro-ph/9601067].

\bibitem{Bennett:2003bz}
C.~L.~Bennett {\it et al.},
``First Year Wilkinson Microwave Anisotropy Probe (WMAP) Observations: Preliminary Maps and Basic Results,''
arXiv:astro-ph/0302207.
\bibitem{Spergel}
D.~N.~Spergel {\it et al.},
``First Year Wilkinson Microwave Anisotropy Probe (WMAP) Observations: Determination of Cosmological Parameters,''
arXiv:astro-ph/0302209.
\bibitem{Peiris}
H.~V.~Peiris {\it et al.},
``First year Wilkinson Microwave Anisotropy Probe (WMAP) observations:  Implications for inflation,''
arXiv:astro-ph/0302225.

\bibitem{toco} A.~D.~Miller {\it et al.},
``A Measurement of the Angular Power Spectrum of the CMB from l = 100 to 400,''
Astrophys.\ J.\  {\bf 524}, L1 (1999)
[arXiv:astro-ph/9906421].

\bibitem{boom}
J.~E.~Ruhl {\it et al.},
``Improved measurement of the angular power spectrum of temperature  anisotropy in the CMB from two new analysis of BOOMERANG observations,''
arXiv:astro-ph/0212229.

\bibitem{maxima}
R.~Stompor {\it et al.}, ``Cosmological implications of the MAXIMA-I
high resolution Cosmic Microwave Background anisotropy measurement,''
Astrophys.\ J.\ {\bf 561}, L7 (2001)
[arXiv:astro-ph/0105062].

\bibitem{dasi} N.~W.~Halverson {\it et al.},
``{DASI} First Results: A Measurement of the Cosmic Microwave Background Angular Power Spectrum,''
Astrophys.\ J.\ {\bf 568}, 38 (2002)
[arXiv:astro-ph/0104489].

\bibitem{cbi}
T.~J.~Pearson {\it et al.},
``The Anisotropy of the Microwave Background to l = 3500: Mosaic Observations with the Cosmic Background Imager,''
arXiv:astro-ph/0205388.
\bibitem{vsa}
K.~Grainge {\it et al.},
``The CMB power spectrum out to l=1400 measured by the VSA,''
arXiv:astro-ph/0212495.
\bibitem{acbar}
C.~l.~Kuo {\it et al.},
``High Resolution Observations of the CMB Power Spectrum with ACBAR,''
arXiv:astro-ph/0212289.


\bibitem{Starobinsky:te}
A.~A.~Starobinsky,
``A New Type Of Isotropic Cosmological Models Without Singularity,''
Phys.\ Lett.\ B {\bf 91}, 99 (1980).

\bibitem{New} A.~D.~Linde,
``A New Inflationary Universe Scenario: A Possible Solution Of The Horizon, Flatness,
 Homogeneity, Isotropy And Primordial Monopole Problems,''
Phys.\ Lett.\ B {\bf 108}, 389 (1982);\\
A.~Albrecht and P.~J.~Steinhardt,
``Cosmology For Grand Unified Theories With Radiatively Induced Symmetry Breaking,''
Phys.\ Rev.\ Lett.\  {\bf 48}, 1220 (1982).
\bibitem{Chaotic}  A.~D.~Linde,
``Chaotic Inflation,''
Phys.\ Lett.\ B {\bf 129}, 177 (1983).

\bibitem{Mukh} V.~F.~Mukhanov and G.~V.~Chibisov,
``Quantum Fluctuation And `Nonsingular' Universe,''
JETP Lett.\  {\bf 33}, 532 (1981)
[Pisma Zh.\ Eksp.\ Teor.\ Fiz.\  {\bf 33}, 549 (1981)]; S.~W.~Hawking,
``The Development Of Irregularities In A Single Bubble Inflationary Universe,''
Phys.\ Lett.\ B {\bf 115}, 295 (1982); 
A.~A.~Starobinsky,
``Dynamics Of Phase Transition In The New Inflationary Universe Scenario And Generation Of Perturbations,''
Phys.\ Lett.\ B {\bf 117}, 175 (1982); A.~H.~Guth and S.~Y.~Pi,
``Fluctuations In The New Inflationary Universe,''
Phys.\ Rev.\ Lett.\  {\bf 49}, 1110 (1982); J.~M.~Bardeen, P.~J.~Steinhardt and M.~S.~Turner, ``Spontaneous Creation Of Almost Scale - Free Density Perturbations In An Inflationary Universe,''
Phys.\ Rev.\ D {\bf 28}, 679 (1983).

\bibitem{Hybrid}
A.~D.~Linde,
``Axions in inflationary cosmology,''
Phys.\ Lett.\ B {\bf 259}, 38 (1991);
A.~D.~Linde,
``Hybrid inflation,''
Phys.\ Rev.\ D {\bf 49}, 748 (1994)
astro-ph/9307002.

\bibitem{Contaldi} C.~R. Contaldi, H. Hoekstra, A. Lewis, 
`` Joint CMB and Weak Lensing Analysis; Physically Motivated
Constraints on Cosmological Parameters,''
arXiv:astro-ph/0302435.

\bibitem{note} For example, compare the interpretation given in the
conclusions of the first two version of~\cite{cline} with the correct
interpretation reported in their third version.

\bibitem{bjk}
J.~R.~Bond, A.~H.~Jaffe and L.~Knox,
``Estimating The Power Spectrum Of The Cosmic Microwave Background,''
Phys.\ Rev.\ D {\bf 57}, 2117 (1998)
[arXiv:astro-ph/9708203].

\bibitem{gaz}
E.~Gaztanaga, J.~Wagg, T.~Multamaki, A.~Montana and D.~H.~Hughes,
``2-point anisotropies in WMAP and the Cosmic Quadrupole,''
arXiv:astro-ph/0304178.

\bibitem{jingfang}
Y.~P.~Jing and L.~Z.~Fang,
``An infrared cutoff revealed by the two years of COBE-DMR
observations of cosmic temperature fluctuations,''
Phys.\ Rev.\ Lett.\ {\bf 73}, 1882 (1994)
[arXiv:astro-ph/9409072].
%
\bibitem{Tegmark:2003ve}
M.~Tegmark, A.~de Oliveira-Costa and A.~Hamilton,
``A high resolution foreground cleaned CMB map from WMAP,''
arXiv:astro-ph/0302496.
\bibitem{Uzan:2003nk}
J.~P.~Uzan, U.~Kirchner and G.~F.~Ellis,
``WMAP data and the curvature of space,''
arXiv:astro-ph/0302597.
\bibitem{Efstathiou}
G. Efstathiou, 
``Is the Low CMB Quadrupole a Signature of Spatial Curvature?''
arXiv:astro-ph/0303127.
\bibitem{Lewis}
S. L. Bridle, A. M. Lewis, J. Weller, G. Efstathiou,
``Reconstructing the primordial power spectrum,'' 
arXiv:astro-ph/0302306.
\bibitem{Linde:2003hc}
A.~Linde,
 ``Can we have inflation with $\Omega > 1$?,''
arXiv:astro-ph/0303245.



\bibitem{Kofman:1986wm}
 L.~A.~Kofman and A.~D.~Linde,
``Generation Of Density Perturbations In The Inflationary Cosmology,''
Nucl.\ Phys.\ B {\bf 282}, 555 (1987); 
L.~A.~Kofman and D.~Y.~Pogosian,
``Nonflat Perturbations In Inflationary Cosmology,''
Phys.\ Lett.\ B {\bf 214}, 508 (1988).



\bibitem{Salopek:1988qh}
D.~S.~Salopek, J.~R.~Bond and J.~M.~Bardeen,
``Designing Density Fluctuation Spectra In Inflation,''
Phys.\ Rev.\ D {\bf 40}, 1753 (1989).

\bibitem{Kofman:1989ed} 
L.~Kofman, G.~R.~Blumenthal, H.~Hodges and J.~R.~Primack,
``Generation Of Nonflat And Nongaussian Perturbations From Inflation,''
ASP Conf.\ Ser.\  {\bf 15}, 339 (1991).

\bibitem{Lidsey:1995np}
J.~E.~Lidsey, A.~R.~Liddle, E.~W.~Kolb, E.~J.~Copeland, T.~Barreiro and M.~Abney,
``Reconstructing the inflaton potential: An overview,''
Rev.\ Mod.\ Phys.\  {\bf 69}, 373 (1997)
arXiv:astro-ph/9508078.

\bibitem{Hodges:bf}
H.~M.~Hodges and G.~R.~Blumenthal,
``Arbitrariness Of Inflationary Fluctuation Spectra,''
Phys.\ Rev.\ D {\bf 42}, 3329 (1990).

\bibitem{Arkani-Hamed:2002fu}
N.~Arkani-Hamed, S.~Dimopoulos, G.~Dvali and G.~Gabadadze,
``Non-local modification of gravity and the cosmological constant  problem,''
arXiv:hep-th/0209227.
N.~Arkani-Hamed, S.~Dimopoulos, G.~Dvali, G.~Gabadadze, and A.D. Linde,
``Self-terminating inflation,''
in preparation.

\bibitem{Linde:1994gy}
A.~D.~Linde, D.~A.~Linde and A.~Mezhlumian,
``Do we live in the center of the world?,''
Phys.\ Lett.\ B {\bf 345}, 203 (1995)
[arXiv:hep-th/9411111].
A.~D.~Linde, D.~A.~Linde and A.~Mezhlumian,
``Nonperturbative Amplifications of Inhomogeneities in a Self-Reproducing Universe,''
Phys.\ Rev.\ D {\bf 54}, 2504 (1996)
[arXiv:gr-qc/9601005].

\bibitem{KS85}
L.~Kofman and A.~A.~Starobinsky,
``Effect Of The Cosmological Constant On Large Scale Anisotropies In The Microwave Backbround,''
Astron.\ Lett.\  {\bf 11}, 271 (1985)
[Pisma Astron.\ Zh.\  {\bf 11}, 643 (1985)].

\bibitem{isw}

L.~R.~Abramo and F.~Finelli,
``Attractors and isocurvature perturbations in quintessence models,''
Phys.\ Rev.\ D {\bf 64}, 083513 (2001)
[arXiv:astro-ph/0101014];
R.~Bean,
``Perturbation evolution with a non-minimally coupled scalar field,''
Phys.\ Rev.\ D {\bf 64}, 123516 (2001)
[arXiv:astro-ph/0104464];
P.~S.~Corasaniti, B.~A.~Bassett, C.~Ungarelli and E.~J.~Copeland,
``Model-independent dark energy differentiation with the ISW effect,''
Phys.\ Rev.\ Lett.\  {\bf 90}, 091303 (2003)
[arXiv:astro-ph/0210209];

\bibitem{Starobinsky:ts}
A.~A.~Starobinsky,
``Spectrum Of Adiabatic Perturbations In The Universe When There Are Singularities In 
JETP Lett.\  {\bf 55}, 489 (1992)
[Pisma Zh.\ Eksp.\ Teor.\ Fiz.\  {\bf 55}, 477 (1992)].
\bibitem{Hodges:1989dw}
H.~M.~Hodges, G.~R.~Blumenthal, L.~A.~Kofman and J.~R.~Primack,
``Nonstandard Primordial Fluctuations From A Polynomial Inflaton Potential,''
Nucl.\ Phys.\ B {\bf 335}, 197 (1990).

\bibitem{Leach:2000yw} 
S.~M.~Leach and A.~R.~Liddle,
``Inflationary perturbations near horizon crossing,''
Phys.\ Rev.\ D {\bf 63}, 043508 (2001)
[arXiv:astro-ph/0010082];
S.~M.~Leach, M.~Sasaki, D.~Wands and A.~R.~Liddle,
``Enhancement of superhorizon scale inflationary curvature perturbations,''
Phys.\ Rev.\ D {\bf 64}, 023512 (2001)
[arXiv:astro-ph/0101406].

\bibitem{Copeland:1994vg}
E.~J.~Copeland, A.~R.~Liddle, D.~H.~Lyth, E.~D.~Stewart and D.~Wands,
``False vacuum inflation with Einstein gravity,''
Phys.\ Rev.\ D {\bf 49}, 6410 (1994)
arXiv:astro-ph/9401011; 
G.~R.~Dvali, Q.~Shafi and R.~K.~Schaefer,
``Large Scale Structure And Supersymmetric Inflation Without Fine Tuning,''
Phys.\ Rev.\ Lett.\  {\bf 73}, 1886 (1994)
[arXiv:hep-ph/9406319].


\bibitem{Linde:1997sj}
A.~D.~Linde and A.~Riotto,
``Hybrid inflation in supergravity,''
Phys.\ Rev.\ D {\bf 56}, 1841 (1997)
arXiv:hep-ph/9703209.


\bibitem{Old} A.~H.~Guth,
`The Inflationary Universe: A Possible Solution To The Horizon And Flatness Problems,''
Phys.\ Rev.\ D {\bf 23}, 347 (1981).


\bibitem{NO}
G.~Felder, L.~Kofman and A.~Linde,
``Inflation and preheating in NO models,''
Phys.\ Rev.\  {\bf D60}, 103505 (1999)
arXiv:hep-ph/9903350.
\bibitem{Cline:2003gq}
J.~M.~Cline,
``Mimicking transPlanckian effects in the CMB with conventional physics,''
arXiv:astro-ph/0303407.

\bibitem{mod1} 
G.~Dvali, A.~Gruzinov and M.~Zaldarriaga,
``A new mechanism for generating density perturbations from inflation,''
arXiv:astro-ph/0303591.

\bibitem{mod2} 
L.~Kofman,
``Probing String Theory with Modulated Cosmological Fluctuations,''
arXiv:astro-ph/0303614.

\bibitem{muk}
V.~F.~Mukhanov,
``Quantum Theory Of Gauge Invariant Cosmological Perturbations,''
Sov.\ Phys.\ JETP {\bf 67}, 1297 (1988)
[Zh.\ Eksp.\ Teor.\ Fiz.\  {\bf 94N7}, 1 (1988\ ZETFA,94,1-11.1988)].
\bibitem{mfb}
V.~F.~Mukhanov, H.~A.~Feldman and R.~H.~Brandenberger,
``Theory Of Cosmological Perturbations,''
Phys.\ Rept.\  {\bf 215}, 203 (1992).

\bibitem{sahni}
V.~Sahni,
``The Energy Density Of Relic Gravity Waves From Inflation,''
Phys.\ Rev.\ D {\bf 42}, 453 (1990).

\bibitem{cline}
J.~M.~Cline, P.~Crotty and J.~Lesgourgues,
``Does the small CMB quadrupole moment suggest new physics?,''
arXiv:astro-ph/0304558.

\bibitem{cmbfast}
U.~Seljak and M.~Zaldarriaga,
``A Line of Sight Approach to Cosmic Microwave Background Anisotropies,''
Astrophys.\ J.\  {\bf 469}, 437 (1996)
arXiv:astro-ph/9603033.
\bibitem{Verde}
L.~Verde {\it et al.},
``First Year Wilkinson Microwave Anisotropy Probe (WMAP) Observations: Parameter Estimation Methodology,''
arXiv:astro-ph/0302218.



\end{thebibliography}
\end{document}